\documentclass[a4paper]{article}
\usepackage{url}
\usepackage{INTERSPEECH2022}
\usepackage{makecell}
\title{Improving Language Identification of Accented Speech}
\name{Kunnar Kukk, Tanel Alumäe}
\address{Laboratory of Language Technology\\Tallinn University of Technology, Estonia}
\email{\{kunkuk, tanel.alumae\}@taltech.ee}

\begin{document}
\maketitle
\begin{abstract}
Language identification from speech is a common preprocessing step in many spoken language processing systems. 
In recent years, this field has seen fast progress, mostly due to the use of self-supervised
models pretrained on multilingual data and the use of large training corpora. 
This paper shows that for speech with a non-native or regional accent, the accuracy of spoken language identification systems drops dramatically, and that the accuracy of identifying the language is inversely correlated with the strength of the accent. We also show that using the output of a lexicon-free speech recognition system of the particular language helps to improve language identification performance on accented speech by a large margin, without sacrificing accuracy  on native speech. We obtain relative error rate reductions ranging from to 35 to 63\% over the state-of-the-art  model across several non-native speech datasets.

\end{abstract}

\section{Introduction}

Spoken language identification (LID) is the task of automatically identifying the language of an utterance. LID is often used as a preprocessing step in speech-based multilingual applications, such as spoken translation, human-machine communication systems and multilingual speech transcription systems. LID is also commonly used in automatic call routing where it is used to direct a call to a fluent native operator \cite{lid_phonexia}. 

Since LID and ASR are increasingly used in critical services, it is important that such systems work flawlessly across a wide user community, 
with respect to variability corresponding to societally sensitive characteristics or traditionally marginalized communities, such as gender, ethnicity,  disability, etc. Decreased robustness of speech based systems towards certain user groups may amplify biases already present in the society.

Speaking in a language other than one’s native tongue is an ubiquitous reality in the globalized world. For example, in 2019, the number of international migrants was estimated to be 272 million, corresponding to 3.5\% of the world’s population \cite{wmt_2020}. The worldwide increase in the number of non-native speakers is caused both by rise in movement due to labour, study and leisure, but also by the large regional conflicts that cause a sudden increase in the  number of refugees in certain parts of the world. Learning the local language is considered one of the most important aspects for migrants’ inclusion in the society by both the receiving community and migrants themselves \cite{castles2002}. A large proportion of migrants are eager to learn the language of the receiving society at least to some degree, as it is important for helping migrants navigate a new environment, including access to health care, banking and other critical services. It also improves their access to education and employment \cite{chiswick2016tongue}. However, almost 90\% of the first generation and around 50\% of the second generation migrants speak the local language with a weak or strong non-native accent \cite{kogan2021ear}. Accent is not a phenomena that is specific to non-native speakers. Also native speakers can have a strong accent peculiar to a particular location or ethnicity, that is different from what is often regarded as a standard pronunciation. However, according to sociolinguistic approach, everyone has an accent, even the native speakers \cite{matsuda1991}. 

In recent years, the field of language identification from speech has seen a fast progress. This is mostly due to the use of self-supervised models trained on very large multilingual datasets (such as XLS-R \cite{babu2021xlsr}) and the emergence of large multilingual speech datasets with language identification labels, such as Mozilla CommonVoice \cite{ardila2019common} and VoxLingua107 \cite{valk_2021}. The resulting models can achieve very high language identification accuracies on spoken data that  contain mostly native speech. The recent Oriental Language Recognition 2021 Challenge \cite{wang2021olr} included a 17 language identification task with utterances obtained from  real-life environments. The top performing teams \cite{lyuant2021,alumae2022} achieved equal error rates below 1\%. This might suggest that LID is a task that is close to be solved.

Several studies have shown that ASR systems produce more errors for non-native speech than for native speech \cite{feng2021quantifying, wu2020see,awasthi2021error}. This is not surprising, since ASR systems are usually trained on speech originating mostly from native speakers.  In \cite{wanneroy2000acoustic}, it was discovered that non-native accent causes on the average three times more LID errors than native speech, when using phonotactic models.

The first goal of this paper is to quantify the accuracy of LID on no-native speech with state-of-the-art models that produce excellent results on native speech. We do this by measuring the performance of different LID models on datasets that contain non-native speech and compare the results with similar datasets containing native speech of the same language. We show that LID systems that provide excellent accuracy for native speech can degrade dramatically in the presence of non-native and regional accents. Then, we investigate using recognition hypotheses of one or many lexicon-free  ASR systems as additional features when producing the LID decision. The idea of using ASR hypotheses for improving LID systems is not entirely new: both  \cite{wang2019signal} and \cite{chandak2020streaming} experimented with combining acoustic and ASR-based features for improving LID and report around 50\% relative reduction in error rate over the baseline acoustic model.
We show that combining  character n-gram based Naïve Bayes text classification models with a system that uses acoustic representations increases the robustness of LID systems to accented speech by a large margin, without sacrificing accuracy on native speech.


\section{Experimental set-up}

\subsection{Datasets}

The following section gives an overview of the 6 datasets used in this work. Table \ref{tab:dataset_statistics} summarises characteristics of datasets.

\begin{table*}[ht!]
    \caption{Characteristics of datasets used for training and evaluation.}
    \label{tab:dataset_statistics}
    \centering
    \begin{tabular}{l c c r c r r} 
    \toprule
         \makecell[l]{\textbf{\thead{Dataset}}} & 
         \textbf{\thead{Language}} & 
         \textbf{\thead{Non-native\\accent?}} & 
         \textbf{\thead{Sampling \\ Rate (kHz)}} & 
         \textbf{\thead{Type}} & 
         \textbf{\thead{Utterances}} & 
         \textbf{\thead{Utterance Avg \\ Length (sec)}} \\
    \midrule
        Estonian Foreign Accent Corpus & Estonian & Yes/No & 44.1 & Spontaneous/Dictated & 32649 & 5.9 \\
        CSLU Foreign Accented English & English & Yes & 8 & Spontaneous & 4925 & 17.9 \\
        CSLU 22 Languages (English) & English & No & 8 & Spontaneous/Dictated & 2206 & 6.4 \\
        CMU Arctic & English & No & 16 & Dictated & 14471 & 3.2 \\
        L2 Arctic & English & Yes & 44.1 & Dictated & 25758 & 3.7 \\
        VoxLingua107 train & 107 & No & 16 & Spontaneous & 2.54M & 9.4 \\
        VoxLingua107 dev & 33 & No & 16 & Spontaneous & 1608 & 10.0 \\ 
    \bottomrule
    \addlinespace
    \textsuperscript{}
    \end{tabular}
\end{table*}

\subsubsection{Estonian Foreign Accent Corpus}
\begin{figure}[ht!]
  \includegraphics[width=\linewidth]{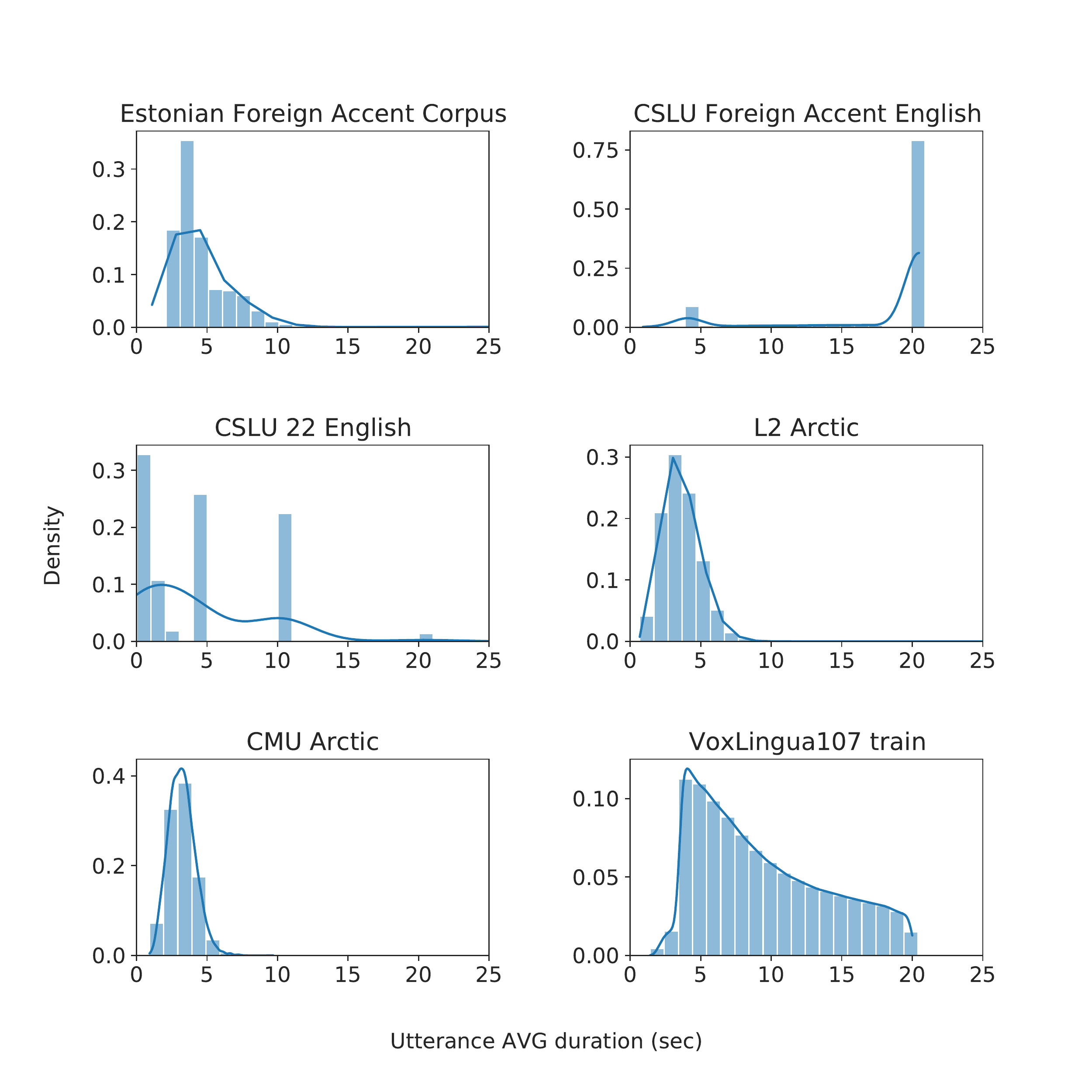}
  \caption{Density of dataset utterance average duration.}
  \label{fig:utt-avg-len-density}
\end{figure}

Estonian Foreign Accent Corpus (EFAC,  version 1) consists of speech data from 185 non-native (L2) and 20 native Estonian speakers. It contains 25-30 minutes of speech from each speaker \cite{meister2012aktsendikorpus, ut_metashare_efac}. Speech is recorded in a studio, using is 16-bit 44.1 kHz stereo format. The dataset consists of 32649 utterances, totalling in 53.2h of speech (48.8h non-native and 3.4h native speech)  with average utterance length of 5.9 seconds. Figure \ref{fig:utt-avg-len-density} shows distribution of utterance lengths of each dataset. 

EFAC speech corpus contains examples of spontaneous and read speech (136 phonetically rich sentences and two short texts). The text corpus involves 130 neutral sentences including the main phonological oppositions of Estonian, eight questions, two passages, and prompts to elicit spontaneous speech (self-introduction, description of three pictures). The dataset also contains the subjects' self-assessment with regard to their Estonian proficiency level.

\subsubsection{CSLU Foreign Accented English}
CSLU Foreign Accented English (CSLU FAE) release 1.2 \cite{cslu_fae_web} consists of non-native speech in English by native speakers of 22 different languages.
Speech is  recorded via a telephone channel, using 16-bit 8kHz mono format. It includes spontaneous telephone speech, information about the speakers' linguistic backgrounds and perceptual judgments about the accents in the utterances. The speakers were asked to speak about themselves in English for 20 seconds, having an average utterance length of 17.9 seconds with a total 24 hours of 4925 telephone-quality utterances. 

\subsubsection{CSLU 22 Languages English Subset}

The English subset of the CSLU 22 Languages Corpus \cite{cslu_22_en}   is a 3.9 hour  dataset of native English. It contains speech recorded using 16-bit 8kHz mono format. Utterances contain both fixed vocabulary words as well as fluent continuous telephone speech.  The number of utterances in the dataset is 2206, having an average utterance length of 6.4 seconds. As seen on Figure \ref{fig:utt-avg-len-density}, a large proportion of the utterances are shorter than one second.

\subsubsection{CMU Arctic}
CMU Arctic dataset \cite{Kominek_cmu_arctic_03} consists of 12.9 hours of dictated speech from 18 native English speakers with various American accents as well as Canadian, Scottish and Indian accents.  It contains 16-bit  16kHz mono-formatted data. 	
Total number of utterances is 14471	with average utterance length of 3.2 seconds. 

\subsubsection{L2 Arctic}

L2 Arctic dataset \cite{l2_arctic_web} consists of 26.4 hours of dictated English speech from 24 non-native speakers with 18 different language backgrounds, with an average of 67.7 minutes of speech per speaker. Speech is recorded using 16-bit 44.1 kHz mono format.The total number of utterances is  25758, with an average utterance length of 3.7 sec. 

\subsubsection{VoxLingua107}

VoxLingua107 train set consists of 6628 hours of speech extracted from automatically scraped Youtube videos \cite{valk_2021}. The language labels of the utterances are based on the detected language of the title and description of the video.
Data-driven post-filtering was used to remove segments from the database that were likely not in the given language, increasing the proportion of correctly labeled segments in the dataset to 98\%, based on crowd-sourced verification. VoxLingua107 has speech data across 107 different languages. Number of utterances in the dataset is about 2.54 million. The average utterance length is 9.4 seconds. The average amount of data per language is 62 hours. 

VoxLingua107 development set consists of 4.5 hours of spontaneous speech from Youtube videos. It has speech data across 33 different languages. The language of the utterances in the development set has been verified by at least two native or proficient crowd-sourced speakers. Number of utterances in the dataset is 1608.

\subsection{Methodology}

In order to assess the impact of  foreign accent on LID performance, we train several LID models on the VoxLingua107 training data and test their accuracy on the English and Estonian test sets and the VoxLingua107 development set. Our first goal is to find out how well current state-of-the-art LID models perform on non-native speech. Our second goal is to propose methods for improving the accuracy of LID on foreign-accented speech, without using any additional training data nor changing the priors of the model, and also without reducing the accuracy of LID on native speech. 

When analyzing the accuracy of different LID models across the presented datasets, it is important to understand that there are many factors besides accent that impact the results, such as the length of the utterance, noise level and type of speech. The native and non-native subsets in EFAC are directly comparable, since they originate from the same dataset. Similarly, the data in CMU Arctic is very similar most aspects to that of L2 Arctic. However, the data in CSLU Foreign Accented English is quite different from the CSLU 22 Languages corpus. Although they both contain telephone speech, a large proportion of the utterances in the CSLU Foreign Accent English corpus are around 20 seconds in length, while the utterances in CSLU 22 English corpus are shorter, with many less than one second in length, which is expected to be very challenging for LID models.

\subsection{Models}

\begin{table*}[tb]
    \caption{Language identification accuracy of different models across the English  and Estonian  test sets.}
    \label{tab:training_results}
    \centering
    \setlength\tabcolsep{4pt}     
    \begin{tabular}{p{0.2cm} p{4.0cm} | c c c c |c c |c}
    \toprule
    & & \multicolumn{4}{c|}{\textbf{English}} & \multicolumn{2}{c|}{\textbf{Estonian}} & \textbf{Various} \\
          &  
         &
         \textbf{CMU Arctic} &
         \textbf{L2 Arctic} &
         \textbf{CSLU FAE} &
         \textbf{CSLU 22 en} &
         \textbf{EFAC}  &
         \textbf{EFAC}  &
        \textbf{ V107 dev}
         \\
\textbf{ID} & \textbf{Model} & \textbf{Native} &\textbf{ Non-native} & \textbf{Non-native} & \textbf{Native} & \textbf{Native} &\textbf{ Non-native} & \textbf{Native}\\         
    \midrule
A & Resnet & 77.4 & 60.5 & 67.1 & 57.7 & 93.3 & 43.9 & 91.9 \\
B & XLS-R 300M & 87.6 & 74.6 & 79.5 & 71.9 & \textbf{99.6} & 51.8 & \textbf{95.3} \\ 
\midrule
D & NB on en ASR  char 4-grams & 83.8 & 79.8 & 84.7 & 57.1 & 20.2 & 21.0 & 54.6 \\
E & NB on et ASR  char 4-grams & 81.5 & 74.6 & 45.7 & 34.3 & 71.0 & 65.3 & 48.7 \\
F & Fusion of D, E    & 91.9 & 88.2 & 81.7 & 58.7 & 68.2 & 62.8 & 58.5 \\
G & LDA+PLDA on log probs of F & 90.1 & 86.0 & 83.4 & 53.8 & 69.2 & 63.7 & 75.3 \\
H & ConvNet on et+en ASR output & 85.7 & 81.5 & 86.9 & 46.8 & 56.2 & 50.7 & 71.4 \\
\midrule
I & Fusion of B, G & \textbf{95.5} & \textbf{90.5} & \textbf{88.2} & \textbf{72.6} & 99.5 & \textbf{69.5} & \textbf{95.3} \\
\bottomrule
    \end{tabular}
\end{table*}

\subsubsection{Resnet model}

The Resnet-style model is derived from the x-vector paradigm \cite{snyder2018x,snyder2018spoken}, with several enhancements. 
For frame-level feature extraction, we use the Resnet34 \cite{cai2018exploring, he2016deep} architecture where the basic convolutional blocks with residual connections are replaced with squeeze-and-excitation modules \cite{hu2018squeeze,zhou2019deep}. 
The statistics pooling layer that maps frame-level features to segment level features is replaced in our model with a multi-head attention layer \cite{bahdanau2014neural, zhu2018self}. The utterance-level features resulting from the attention-based statstics pooling layer are further processed by two fully connected layers that also apply batch normalization and the ReLU non-linearity. The model is trained using cross-entropy loss. The details of this model can be found in \cite{alumae2022}.

For LID, this model is not applied directly but it is used for extracting the embeddings for the training and test data. Embeddings are extracted from the output of the first fully connected layer that comes after pooling. The embeddings are centered on the training data and reduced to 108-dimensional vectors using Linear Discriminant Analysis (LDA). The final scoring is done using a Probablistic Linear Discriminant Analysis (PLDA) model.

\subsubsection{XLS-R 300M}
We also experimented with using the XLS-R-300M wav2vec2.0 model  \cite{babu2021xlsr} as the backbone of our language embedding model. XLS-R-300M is trained on unlabeled multilingual data. The model is trained by jointly solving a contrastive task over masked latent speech representations  and learning a quantization of the latents shared across languages.  
XLS-R is pretrained on  around 500K hours of speech data from 128 languages. 

We used XLS-R-300M as follows: the outputs from the wav2vec2 model were fed through an attentive pooling  layer, a fully connected layer with ReLU and batch normalization, and the final output layer, corresponding to the languages of the training set. During training, the learning rate corresponding to the XLS-R model was set to 0.01 times lower than the base learning rate. As the final classification backend, similar LDA/PLDA based setup as for the Resnet model was applied. This model achieves state-of-the art results on the VoxLingua107 development set. It was also the main component of the system that was ranked 2nd in the unconstrained task of the OLR 2021 Challenge \cite{alumae2022}.

\subsubsection{Multinominal Naïve Bayes model on ASR output}


The multinomial Naïve Bayes (NB) model predicts the language of an utterance based on its ASR-based transcript. It uses word-internal character 4-grams as features, with n-grams at the edges of words padded with space. The model considers all n-grams that occur in training data and uses Laplace smoothing with the smoothing parameter set to 0.95.

For generating ASR transcripts, we used two models: English and Estonian. Both models are finetuned from the multilingual wav2vec2 models using connectionist temporal classification (CTC) loss. The English model\footnote{\url{https://huggingface.co/jonatasgrosman/wav2vec2-large-xlsr-53-english}} is finetuned from XLSR-53K \cite{conneau2020unsupervised} using the English CommonVoice data. The Estonian model\footnote{\url{https://huggingface.co/TalTechNLP/xls-r-300m-et}} is finetuned from XLS-R-300M \cite{babu2021xlsr} using around 800 hours of diverse Estonian speech (mainly broadcast speech). Neither of those models use an external language model during decoding and  both are using character-based vocabularies. This has two benefits: first, decoding using a GPU is very fast, making it feasible to decode the whole 6628 hours of VoxLingua107 data for generating training data. Second, the output of the lexicon-free ASR system is not constrained to in-vocabulary words, resulting in very expressive character ASR-transcripts for languages other than the ASR target language.

\subsubsection{Convolutional Neural Network on ASR outputs}

As an alternative to a NB-based text classification model, we experimented with a convolutional neural network (ConvNet). The proposed ConvNet consists of several parallel convolutional input branches that each process the ASR transcript generated using a particular ASR model. The outputs from convolutional input branches are pooled over the utterance using max-pooling, concatenated and further processed using two fully connected layers. The model is trained using cross-entropy loss. The convolutional branches first map characters to their learned 20-dimensional embeddings and then apply a series of convolutional layers. In our experiments, we used five 1D convolutional layers with kernel sizes $(3, 1, 3, 1)$, with the number of channels set to 512. 
Similarly to the acoustic-based LID models, the ConvNet model is not applied directly for inference but is used for extracting 512-dimensional embeddings (from the output of the first fully connected layer that comes after pooling). The embeddings are then processed using the LDA/PLDA model.

\section{Results and analysis}


\begin{figure}[t]
  \centering
  \includegraphics[width=0.9\linewidth]{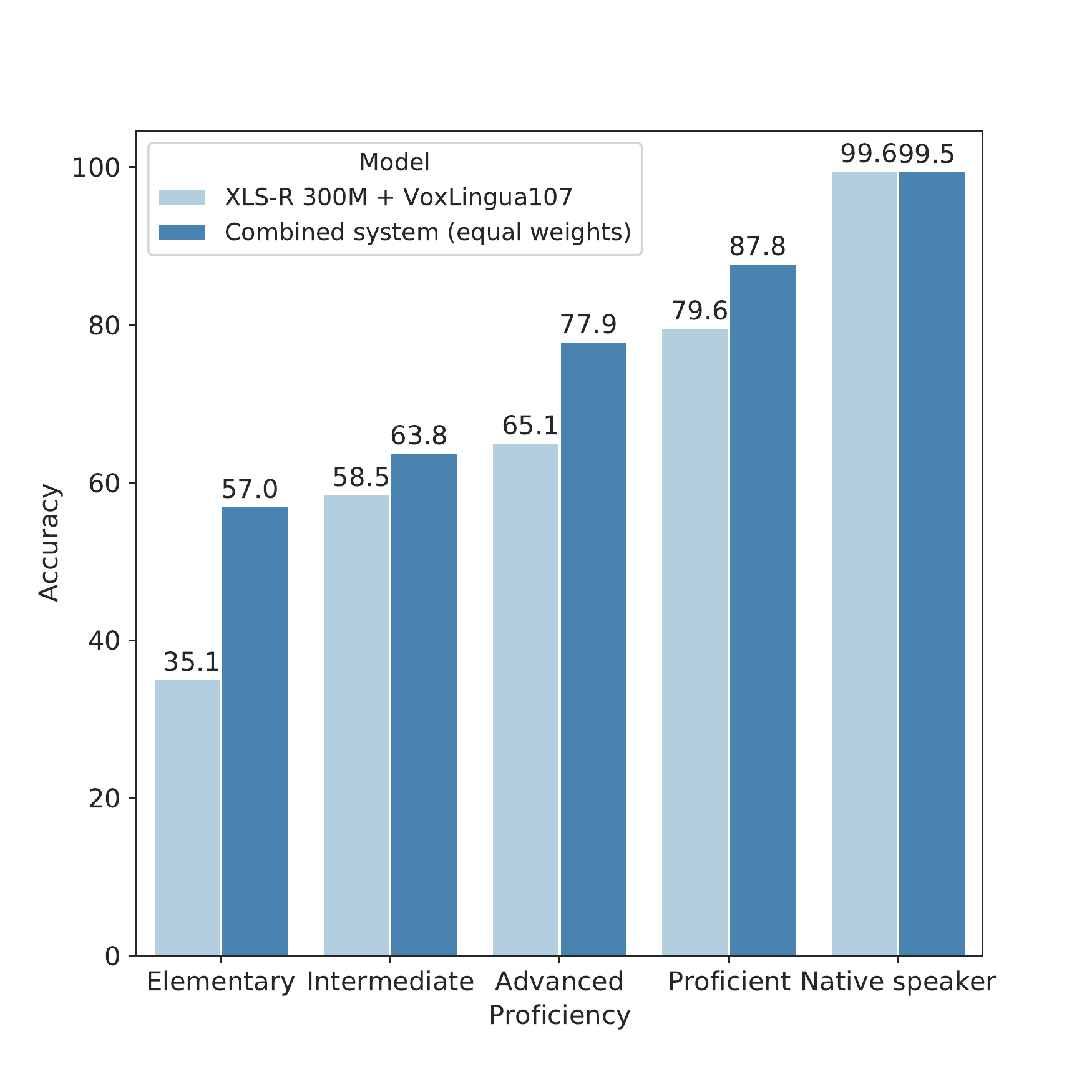}
  \caption{Accuracy of identifying Estonian, depending on the speaker's self-estimated proficiency.}
  \label{fig:est-accent-accuracy}
\end{figure}

The results of different models and their combinations are listed in Table \ref{tab:training_results}. All models are trained on the VoxLingua107 training set.  The first two models (A and B) are based on only acoustic representations. It can be seen that using the finetuned XLS-R wav2vec2 model results in large improvements over the Resnet model on all datasets, regardless of the accent. 
The result on the VoxLingua107 development set is better than the previous state-of-the-art of 94.3\% \cite{babu2021xlsr}.
At the first sight it is surprising that the accuracy on the native English CMU Arctic dataset is much lower than the accuracy on the native Estonian subset of EFAC. Upon deeper inspection, it turns out that the accuracy varies a lot across the different speakers in CMU Arctic, ranging from 35\% for a speaker with a distinctive Indian accent to 100\% for a speaker without any marked pronunciation features (using model B). This indicates that the LID models using acoustic representations not only struggle with non-native speech, but also with native speech with a distinctive regional accent.

Models D-H are all based on ASR transcriptions. By comparing models D (Naïve Bayes model using character 4-grams from transcriptions generated using the English model) and E (same, but using Estonian model transcriptions) it can be seen that having the target language ASR transcripts available is more helpful for LID than other ASR transcripts: e.g., the accuracy of a model trained on the output of the Estonian ASR system (model D) achieves 71\% accuracy on native Estonian, dropping to only 20\% when using English ASR transcripts. Fusion of individual NB models using linear interpolation results in gains in LID performance for all datasets. Using the log posterior probabilities of the fused NB model as input to a LDA/PLDA based LID system results in further improvements for Estonian and VoxLingua107 dev set. Surprisingly, the ConvNet trained on ASR transcripts is not able to outperform the fusion of NB models.

Model I, the fusion of the best acoustic and ASR-based models, outperforms all models on most datasets by a large margin. The fusion uses uniform weights for the two models. For good performance on accented speech it is important not to optimize the fusion weights on  native speech data (such as VoxLingua107): on native speech, acoustic representations result in much higher accuracy than the ASR-based features and the optimized model collapses into an acoustic-only model.

%

Figure \ref{fig:est-accent-accuracy} compares the accuracy of the acoustic model (B) and the fused model (I) on the EFAC data, using subjects' self-estimated proficiency to group the speakers. The chart confirms that there is a strong inverse correlation between the strength of the accent and the LID performance. Using ASR transcripts as additional features improves LID results across all proficiency levels for non-native speech. However, there is still a noticeable gap between LID accuracy of native and non-native speech, even for proficient non-native speakers.

\section{Conclusion}

In this paper, we have shown that LID systems that perform exceptionally well on native speech, have dramatically worse accuracy on identifying the language from non-native speech and native speech with a distinctive regional accent. 

Experiments showed that this problem can be mitigated (but not fully solved) by using a LID model that fuses the predictions of the acoustic-based model with the outputs of a text classification model trained on the transcripts of one or many monolingual lexicon-free ASR systems. In our experiments, this helped to improve LID accuracy on non-native speech by a large margin, with relative error rate reductions ranging from to 35 to 63\% over the state-of-the-art acoustic model, without decreasing accuracy of LID on native speech.

\bibliographystyle{IEEEtran}
\bibliography{mybib}

\end{document}